\documentclass[prd,preprint,tightenlines,showpacs,floatfix,showpacs,preprintnumbers,nofootinbib,eqsecnum,superscriptaddress]{revtex4}

 \usepackage[dvips,final]{graphicx}
  \usepackage{amssymb}
   \usepackage{amsmath}
    \usepackage{amsfonts}
     \usepackage{epsfig}
      \usepackage{bm}
\begin{document}

\title{
Nonphotonic electrons at RHIC \\
within $k_t$-factorization approach\\
and with experimental semileptonic decay functions.
}

\author{M. {\L}uszczak}
\email{luszczak@univ.rzeszow.pl}
\affiliation{University of Rzesz\'ow, PL-35-959 Rzesz\'ow, Poland}

\author{R. Maciu{\l}a}
\email{rafal.maciula@poczta.fm}
\affiliation{University of Rzesz\'ow, PL-35-959 Rzesz\'ow, Poland}

\author{A. Szczurek}
\email{Antoni.Szczurek@ifj.edu.pl}
\affiliation{Institute of Nuclear Physics PAN, PL-31-342 Cracow,
Poland} 
\affiliation{University of Rzesz\'ow, PL-35-959 Rzesz\'ow, Poland}

\date{today}

\begin{abstract}
We discuss production of nonphotonic electrons
in proton-proton scattering at RHIC.
The distributions in rapidity and transverse momentum
of charm and bottom quarks/antiquarks are calculated in 
the $k_t$-factorization approach. We use different
unintegrated gluon distributions from the literature.
The hadronization of heavy quarks is done by means of 
Peterson and Braaten et al. fragmentation functions.
The semileptonic decay functions are found by fitting
recent semileptonic data obtained by the CLEO and BABAR
collaborations.
We get good description of the data at large transverse 
momenta of electrons and find a missing strength 
concentrated at small transverse momenta of electrons.
Plausible missing mechanisms are discussed.
\end{abstract}

\pacs{12.38.-t,12.38.Cy,14.65.Dw}

\maketitle

\section{Introduction}

Recently the PHENIX and STAR collaborations
has measured transverse momentum distribution
of so-called nonphotonic electrons \cite{PHENIX_electrons,
STAR_electrons}.
It is believed that the dominant contribution
to the nonphotonic electrons/positrons comes from the
semileptonic decays of charm and beauty mesons.
These processes have three subsequent stages.
First $c \bar c$ or $b \bar b$ quarks are produced.
The dominant mechanisms being gluon-gluon fusion
and quark-antiquark annihilation.
Next the heavy quarks/antiquarks are turned to heavy
charmed mesons $D, D^*$ or $B, B^*$. The vector
$D^*$ and $B^*$ mesons decay strongly producing 
$D$ and $B$ mesons.
Finally the heavy pseudoscalar mesons decay 
semileptonically producing electrons/positrons.

\begin{figure}[!thb]
\begin{center}
\includegraphics[width=7.0cm]{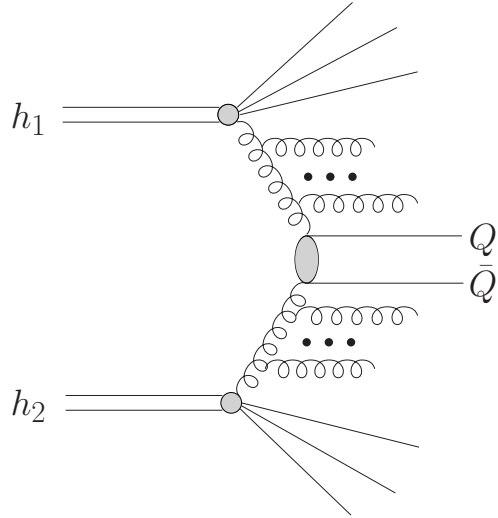}
\caption[*]{
A basic diagram relevant for gluon-gluon fusion
in $k_t$-factorization.
\label{fig:diagram_fusion}
}
\end{center}
\end{figure}

The inclusive heavy quark/antiquark production
can presently be calculated at
Fixed-Order plus Next-to-Leading-Log (FONLL) level
\cite{FONLL}. The predictions for
electron spectra in proton-proton collisions at RHIC 
can be found in Ref.\cite{CNV05}.
An alternative approach for inclusive 
heavy quark production is $k_t$-factorization
\cite{CCH91,CE91,BE01,RSS,BS00,HKSST00,LSZ02}. 
In this approach emission of gluons 
(see Fig.\ref{fig:diagram_fusion}) is encoded in 
so-called unintegrated gluon distributions (UGDFs).
The latter approach is very efficient for description 
of $Q \bar Q$ correlations \cite{LS06}.

The hadronization of heavy quarks is usually done
with the help of fragmentation functions.
The Peterson fragmentation functions are often
used in this context \cite{Peterson}. 
The parameters of the Peterson fragmentation functions 
are adjusted to $e^+ e^-$
or $p \bar p$ production of heavy mesons.
Another perturbative fragmentation model has been 
proposed in Ref.\cite{BCFY95} (BCFY).

The last ingredient are semileptonic decays
of heavy mesons.
Until recently this component was treated by
modeling the decay \cite{Hill,Mahlke07,AMP08}. Only 
recently the CLEO \cite{CLEO} and BABAR \cite{BABAR} 
collaborations has measured very precisely
the spectrum of electrons/positrons coming from
the decays of $D$ and $B$ mesons, respectively.
This is done by producing resonances: $\Psi(3770)$
which decays into $D$ and $\bar D$ mesons (CLEO) and
$\Upsilon(4S)$ which decays into $B$ and $\bar B$ mesons
(BABAR).
In both cases the heavy mesons are almost at rest,
so in practice one measures the meson rest frame
distributions of electrons/positrons.

In the present analysis we shall apply 
the $k_t$-factorization approach. At relatively low
RHIC energies rather intermediate $x$-values
become relevant.  The Kwiecinski unintegrated gluon
(parton) distributions seem relevant in this case
\cite{Kwiecinski}.
We shall use also Ivanov-Nikolaev distributions
which were fitted to deep-inelastic HERA data 
including intermediate-$x$ region \cite{IN02}.
We shall use both Peterson and BCFY fragmentation
functions. The electron/positron decay functions
will be fitted to the recent CLEO and BABAR data.

\section{Formalism}

Let us consider the reaction $h_1 + h_2 \to Q + \bar Q + X$,
where $Q$ and $\bar Q$ are heavy quark and heavy antiquark,
respectively.
In the leading-order (LO) approximation within the collinear approach
the quadruply differential cross section in the rapidity 
of $Q$ ($y_1$),
in the rapidity of $\bar Q$ ($y_2$) and the transverse momentum of
one of them ($p_t$) can be written as
\begin{equation}
\frac{d \sigma}{d y_1 d y_2 d^2p_t} = \frac{1}{16 \pi^2 {\hat s}^2}
\sum_{i,j} x_1 p_i(x_1,\mu^2) \; x_2 p_j(x_2,\mu^2) \;
\overline{|{\cal M}_{ij}|^2} \; .
\label{LO_collinear}
\end{equation}
Above, $p_i(x_1,\mu^2)$ and $p_j(x_2,\mu^2)$ are the familiar
(integrated) parton distributions in hadron $h_1$ and $h_2$, respectively.
There are two types of the LO $2 \to 2$ subprocesses which enter
Eq.(\ref{LO_collinear}): $gg \to Q \bar Q$ and $q \bar q \to Q \bar Q$.
The first mechanism dominates at large energies and the second one
near the threshold. The parton distributions are evaluated at:
$x_1 = \frac{m_t}{\sqrt{s}}\left( \exp( y_1) + \exp( y_2) \right)$,
$x_2 = \frac{m_t}{\sqrt{s}}\left( \exp(-y_1) + \exp(-y_2) \right)$,
where $m_t = \sqrt{p_t^2 + m_Q^2}$.
The formulae for matrix element squared averaged over the initial
and summed over the final spin polarizations can be found e.g. in
Ref.\cite{BP_book}.

If one allows for transverse momenta of the initial partons,
the sum of transverse momenta of the final $Q$ and $\bar Q$ no longer
cancels.
Formula (\ref{LO_collinear}) can be easily generalized if
one allows for the initial parton transverse momenta. Then
\begin{eqnarray}
\frac{d \sigma}{d y_1 d y_2 d^2p_{1,t} d^2p_{2,t}} = \sum_{i,j} \;
\int \frac{d^2 \kappa_{1,t}}{\pi} \frac{d^2 \kappa_{2,t}}{\pi}
\frac{1}{16 \pi^2 (x_1 x_2 s)^2} \; \overline{ | {\cal M}_{ij} |^2}
\nonumber \\  
\delta^{2} \left( \vec{\kappa}_{1,t} + \vec{\kappa}_{2,t} 
                 - \vec{p}_{1,t} - \vec{p}_{2,t} \right) \;
{\cal F}_i(x_1,\kappa_{1,t}^2) \; {\cal F}_j(x_2,\kappa_{2,t}^2) \; ,
\label{LO_kt-factorization}    
\end{eqnarray}
where now ${\cal F}_i(x_1,\kappa_{1,t}^2)$ and ${\cal F}_j(x_2,\kappa_{2,t}^2)$
are the so-called unintegrated gluon (parton) distributions
\footnote{In this paper we shall use the following convention
of unintegrated gluon distributions:
$\int_0^{\mu^2} {\cal F}(x,\kappa^2) d \kappa^2 \sim x g(x,\mu^2)$
}.
The extra integration is over transverse momenta of the initial
partons.
The two extra factors $1/\pi$ attached to the integration over
$d^2 \kappa_{1,t}$ and $d^2 \kappa_{2,t}$ instead over
$d \kappa_{1,t}^2$ and $d \kappa_{2,t}^2$ as in the conventional
relation between the unintegrated ($\cal F$) and the integrated ($g$)
parton distributions. 
The two-dimensional Dirac delta function assures momentum conservation.
Now the unintegrated parton distributions must be evaluated at:
$x_1 = \frac{m_{1,t}}{\sqrt{s}}\exp( y_1) 
     + \frac{m_{2,t}}{\sqrt{s}}\exp( y_2)$,
$x_2 = \frac{m_{1,t}}{\sqrt{s}}\exp(-y_1) 
     + \frac{m_{2,t}}{\sqrt{s}}\exp(-y_2)$,
where $m_{i,t} = \sqrt{p_{i,t}^2 + m_Q^2}$.
In general, the matrix element must be calculated for initial
off-shell partons. The corresponding formulae for initial gluons
were calculated in \cite{CCH91,CE91} (see also \cite{BE01}).
It is easy to check \cite{LS06} that in the limit
$\kappa_1^2 \to 0$, $\kappa_2^2 \to 0$
the off-shell matrix elements converge to
the on-shell ones.

Introducing new variables:
\begin{eqnarray}
\vec{Q}_t = \vec{\kappa}_{1,t} + \vec{\kappa}_{2,t} \; , \nonumber \\
\vec{q}_t = \vec{\kappa}_{1,t} - \vec{\kappa}_{2,t} \; 
\label{new_variables}
\end{eqnarray}
we can write:
\begin{eqnarray}
\frac{d \sigma_{ij}}{d y_1 d y_2 d^2p_{1,t} d^2p_{2,t}} =
\int d^2 q_t \; \frac{1}{4 \pi^2}
\frac{1}{16 \pi^2 (x_1 x_2 s)^2} \; \overline{ | {\cal M}_{ij} |^2}
\nonumber \\  
{\cal F}_i(x_1,\kappa_{1,t}^2) \; {\cal F}_j(x_2,\kappa_{2,t}^2) \; .
\label{LO_kt-factorization2}    
\end{eqnarray}
This formula is very useful to study correlations between
the produced heavy quark $Q$ and heavy antiquark 
$\bar Q$ \cite{LS06}.

For example
\begin{eqnarray}
\frac{d \sigma_{ij}}{d p_{1,t} d p_{2,t}} &=&
\int d \phi_1 d \phi_2 \; p_{1,t} p_{2,t} \int dy_1 d y_2
\int d^2 q_t \; \frac{1}{4 \pi^2}
\frac{1}{16 \pi^2 (x_1 x_2 s)^2} \; \overline{ | {\cal M}_{ij} |^2}
\nonumber \\  
&&{\cal F}_i(x_1,\kappa_{1,t}^2) \; {\cal F}_j(x_2,\kappa_{2,t}^2) 
\nonumber \\
&=& 4 \pi \; \frac{1}{2} \; \frac{1}{2} \;
 \int d \phi_{-} \; p_{1,t} p_{2,t} \int dy_1 d y_2
\int d^2 q_t \; \frac{1}{4 \pi^2}
\frac{1}{16 \pi^2 (x_1 x_2 s)^2} \; \overline{ | {\cal M}_{ij} |^2}
\nonumber \\  
&&{\cal F}_i(x_1,\kappa_{1,t}^2) \; {\cal F}_j(x_2,\kappa_{2,t}^2)  \; .
\label{p1t_p2t_map}
\end{eqnarray}
In the last equation we have introduced $\phi_{-} \equiv \phi_1 -
\phi_2$, where $\phi_{-} \in$ (-2$\pi$, 2$\pi$).
The factor 4 $\pi$ comes from the integration over
$\phi_{+} \equiv \phi_1 + \phi_2$. The first factor 1/2 comes from
the jacobian transformation while the second factor 1/2
takes into account an extra extension of the domain
when using $\phi_{+}$ and $\phi_{-}$ instead of $\phi_{1}$
and $\phi_{2}$.

At the Tevatron and LHC energies the contribution of 
the $gg \to Q \bar Q$ subrocess is more than an order 
of magnitude larger than its counterpart for 
the $q \bar q \to Q \bar Q$ subprocess. At RHIC energy
the relative contribution of the quark-antiquark 
annihilation is somewhat bigger.
Therefore in the following we shall take into account
not only gluon-gluon fusion process i.e. i=0 and j=0
but also the quark-antiquark annihilation mechanism.

The purely perturbative\footnote{when both UGDFs are generated perturbatively}
$k_t$-factorization formalism to $h_1 h_2 \to Q \bar Q$ applies if
$\kappa_{1,t}^2, \kappa_{2,t}^2 > \kappa_0^2$.
The choice of $\kappa_0^2$ is to a large extent arbitrary.
In Refs.\cite{RSS} a rather large $\kappa_0^2$ was chosen
and the space $\kappa_{1,t}^2 \times \kappa_{2,t}^2$ was
subdivided into four disjoint regions. For example
the contribution when both $\kappa_{1,t}^2$ and $\kappa_{2,t}^2$
are small was replaced by the leading-order collinear cross section.
Such an approach assures that
$\sigma_{Q \bar Q}^{tot} > \sigma_{Q \bar Q}^{tot}$(collinear LO)
by construction. \\
It is rather obvious that the resulting cross section 
strongly depends on the choice of $\kappa_0^2$
which makes the procedure a bit arbitrary. 
Our philosophy here is different. Many models of UGDF 
in the literature treat the soft region explicitly.
Therefore we use the $k_t$-factorization formula 
everywhere on the $\kappa_{1,t}^2 \times \kappa_{2,t}^2$ 
plane. 

The production of electrons/positrons is a multi-step
process.
The whole procedure of electron/positron production
can be written in the following schematic way:
\begin{equation}
\frac{d \sigma^e}{d y d^2 p} =
\frac{d \sigma^Q}{d y d^2 p} \otimes
D_{Q \to D} \otimes
f_{D \to e} \; ,
\label{whole_procedure}
\end{equation}
where the symbol $\otimes$ denotes a generic convolution.
The first term responsible for production
of heavy quarks/antiquarks is calculated
in the $k_t$-factorization approach.
Some details were already discussed above.
Next step is the process of formation of heavy mesons.
We follow a phenomenological approach and take Peterson 
and Braaten et al. fragmentation functions
with parameters from the literature (see e.g. \cite{PDG}).
The electron decay function should account for the proper
branching fractions. The latter are known experimentally 
(see e.g. \cite{PDG,CLEO,BABAR}). 
These functions can in principle be calculated 
\cite{Hill,AMP08}. This introduces, however, some model 
uncertainties and requires inclusion of all final 
state channels explicitly. An alternative is to use 
experimental input.
The decay functions have been measured only recently
\cite{CLEO,BABAR}. How to use the recent experimental 
information will be discussed in the next section.

\section{Results}

\begin{figure}[!thb] 
\begin{center}
\includegraphics[width=8.0cm]{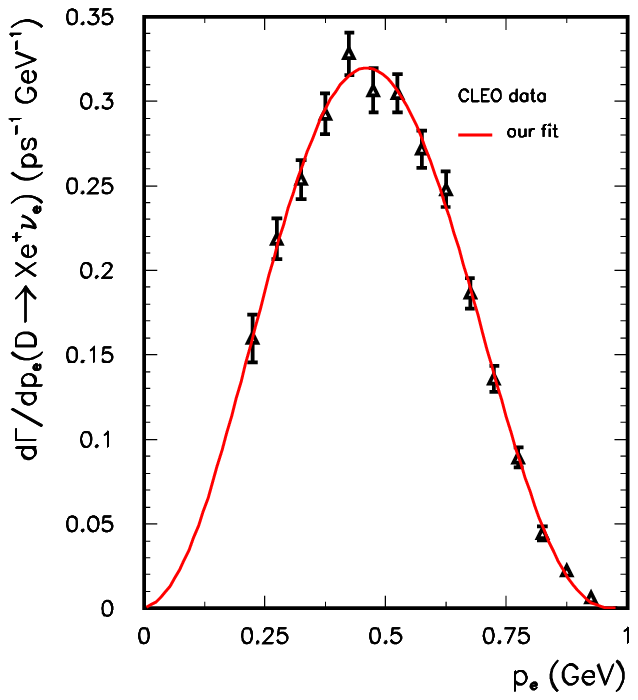}
\includegraphics[width=8.0cm]{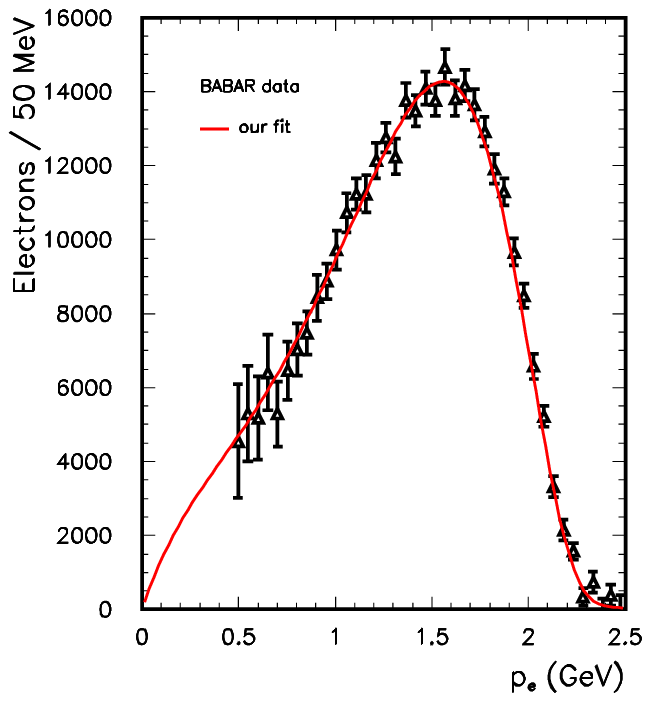}
\caption[*]{
Our fit to the CLEO \cite{CLEO} and BABAR \cite{BABAR}
data.
\label{fig:cleo_and_babar}
}
\end{center}
\end{figure}

In principle, the semileptonic decays can be modeled
(see e.g. \cite{Hill,Mahlke07,AMP08}).
Since there are many decay channels with different
number of particles this is not an easy task.
In our approach we take less ambitious but more pragmatic
approach.
In Fig.\ref{fig:cleo_and_babar} we show our
purely mathematical fit to not absolutely normalized
data of the CLEO \cite{CLEO} and BABAR \cite{BABAR}
collaborations. We find a good fit with:
\begin{equation}
f_{CLEO}(p) = 12.55 (p+0.02)^{2.55} (0.98-p)^{2.75}
\label{CLEO_fit_function}
\end{equation}
for the CLEO data \cite{CLEO} and
\begin{equation}
\begin{split}
f_{BABAR}(p) = \left( 126.16+14293.09 
       \exp(-2.24 \ln(2.51-0.97 p)^2 \right) \\
\left( -41.79+42.78 \exp(-0.5(|p-1.27|) /1.8 )^{8.78} \right )
\end{split}
\label{BABAR_fit_function}
\end{equation}
for the BABAR data \cite{BABAR}.
In these purely numerical parametrizations
$p$ must be taken in GeV.

After renormalizing to experimental branching fractions
for $D \to~e$ (about 10 \% \footnote{The branching fraction
for different species of $D$ mesons is different:
BR$(D^+\to~e^+ \nu_e X)$=16.13$\pm$0.20(stat.)$\pm$0.33(syst.)\%,
BR$(D^0\to~e^+ \nu_e X)$=6.46$\pm$0.17(stat.)$\pm$0.13(syst.)\%) 
\cite{CLEO}. Because the shapes of positron spectra for both decays are 
identical within error bars we can take the average value and 
simplify the calculation.} and 
$B\to e$ (10.36 $\pm$ 0.06(stat.) $\pm$ 0.23(syst.) \% \cite{BABAR} )
we shall use them to generate electrons/positrons 
in the rest frame of the decaying $D$ and $B$ mesons 
in a Monte Carlo approach. 
We shall neglect a small effect of 
the non-zero motion of the $D$ mesons in 
the case of the CLEO experiment and of the $B$ mesons
in the case of the BABAR experiment. This effect
is completely negligible.

\begin{figure}[!thb]
\begin{center}
\includegraphics[width=5.0cm]{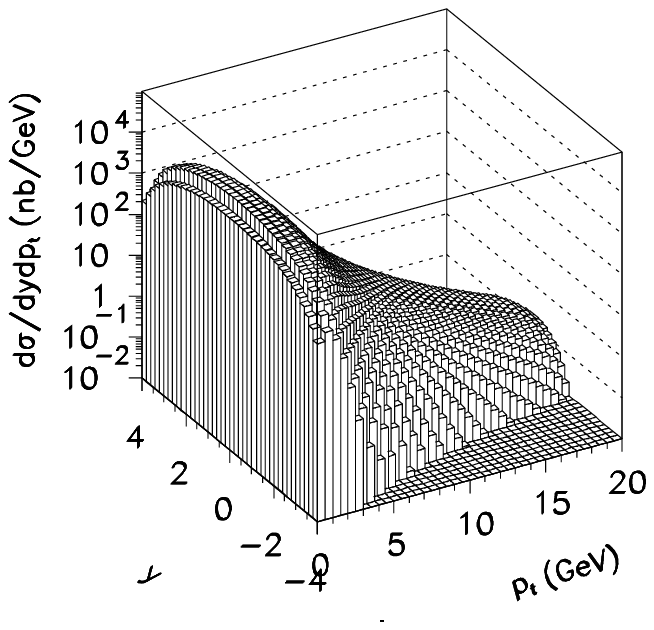}
\includegraphics[width=5.0cm]{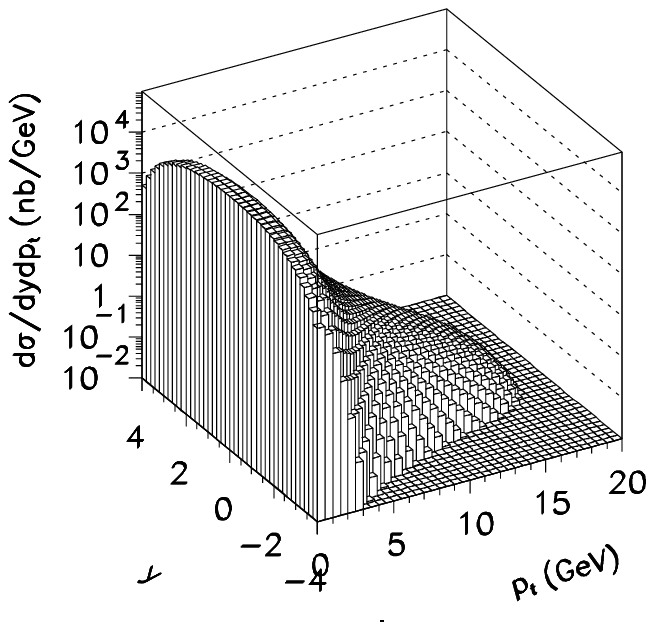}
\includegraphics[width=5.0cm]{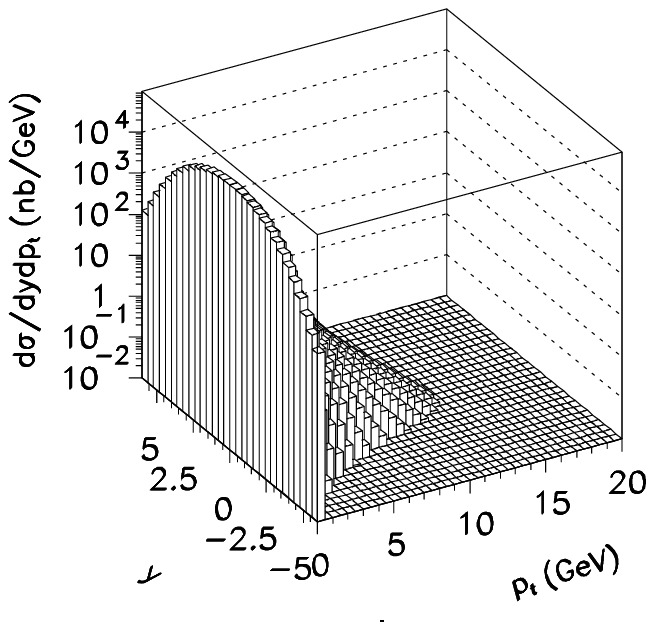}
\caption[*]{
Two-dimensional distributions in rapidity and transverse
momentum for charm quark/antiquark, D mesons and
electrons/positrons.
\label{fig:c_to_D_to_e}
}
\end{center}
\end{figure}

For illustration of the whole procedure in 
Fig.\ref{fig:c_to_D_to_e} we show as an example
two-dimensional distributions in rapidity and 
transverse momentum for charm quarks, $D$ mesons and 
electrons from the decay of $D$ mesons. 
Both fragmentation and semileptonic decays cause 
degradation of transverse momentum. 
On average $p_{t,e} < p_{t,D} < p_{t,c}$. 
The spectra of electrons are much softer
than initial spectra of charm quarks.
On the other hand the distributions in 
rapidity of electrons are much broader than the corresponding 
distributions of quarks/antiquarks.

\begin{figure}[!thb] 
\begin{center}
\includegraphics[width=6cm]{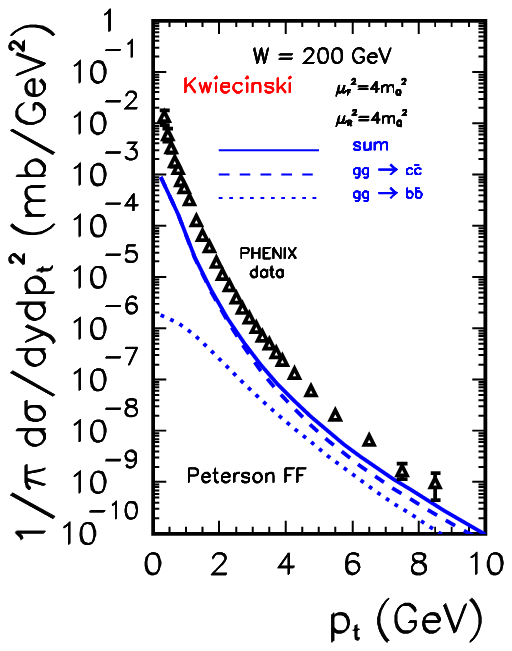}
\includegraphics[width=6cm]{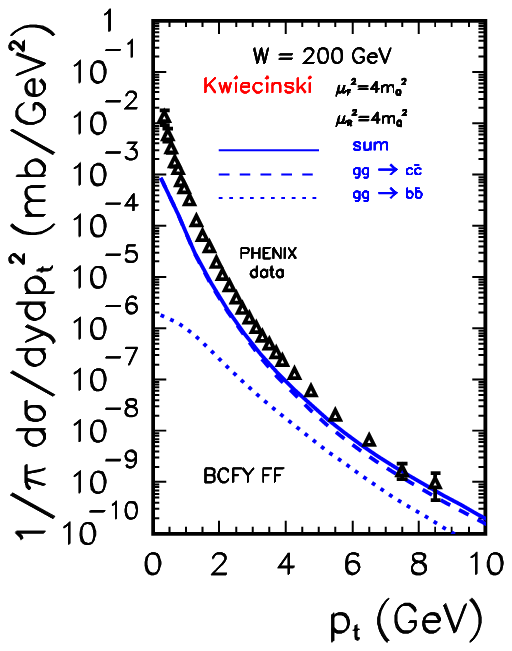}
\includegraphics[width=6cm]{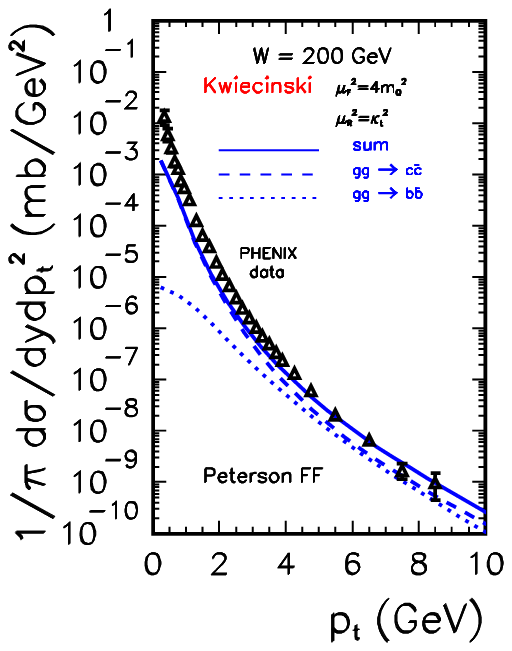}
\includegraphics[width=6cm]{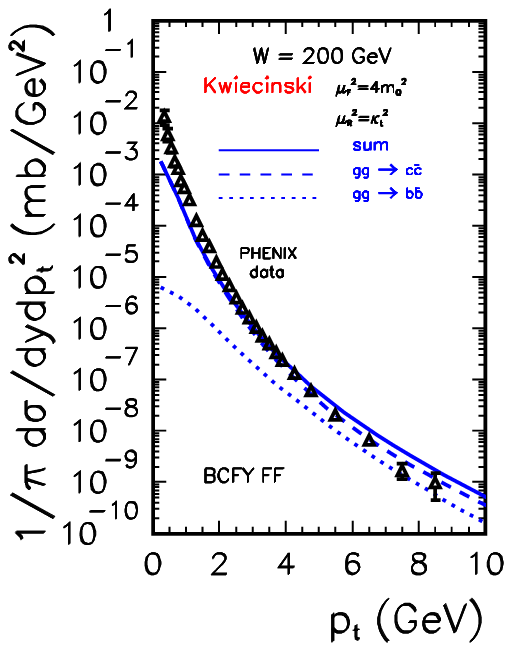}
\caption[*]{
Transverse momentum distribution of electrons/positrons
with the Kwieci\'nski UGDF. Different combinations
of factorization and renormalization scales are used.
On the left side we show results with Peterson
fragmentation functions and on the right side with
BCFY fragmentation functions. 
\label{fig:dsig_dpt_kwiecinski1}
}
\end{center}
\end{figure}

\begin{figure}[!thb] 
\begin{center}
\includegraphics[width=6cm]{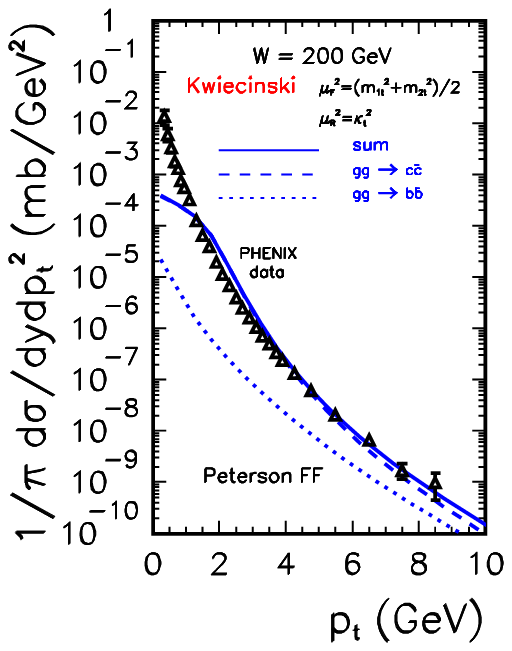}
\includegraphics[width=6cm]{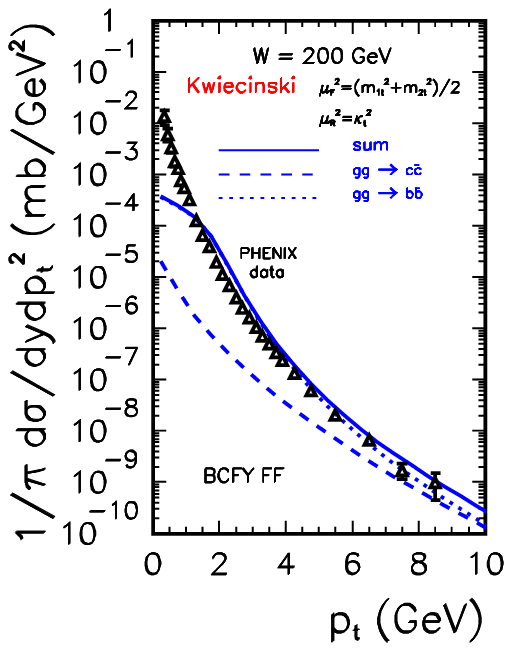}
\includegraphics[width=6cm]{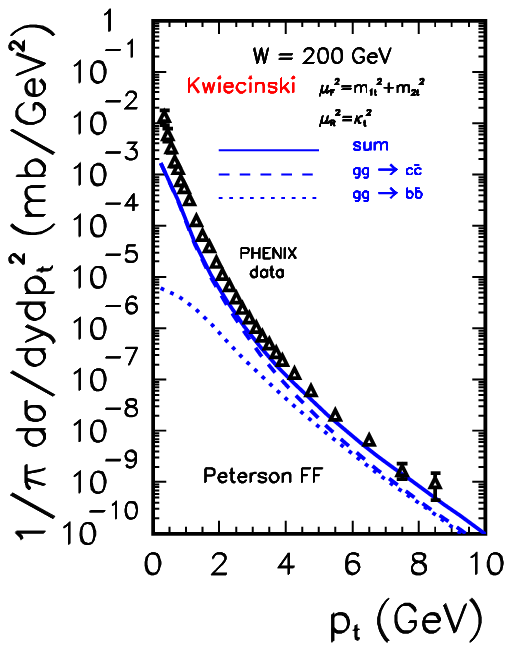}
\includegraphics[width=6cm]{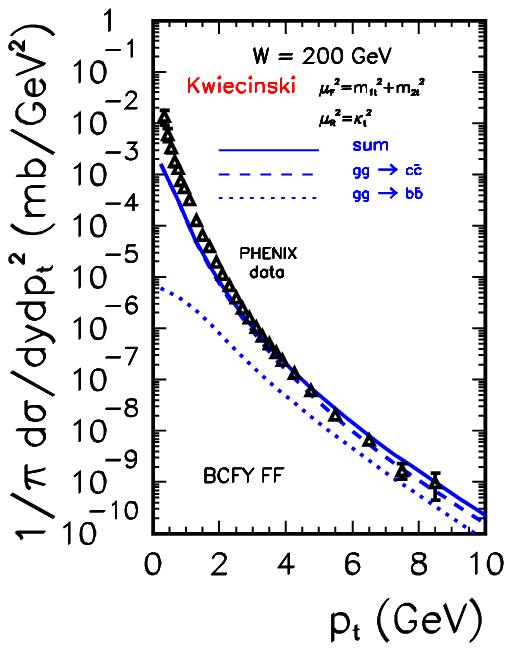}

\caption[*]{
The same as in the previous section but with different choices of factorization/renormalization scale.
\label{fig:dsig_dpt_kwiecinski2}
}
\end{center}
\end{figure}

Now we shall concentrate on invariant cross section
as a function of electron/positron transverse momentum.
Such distributions have been measured recently by
the STAR and PHENIX collaboration at RHIC 
\cite{PHENIX_electrons,STAR_electrons}.
In Fig.\ref{fig:dsig_dpt_kwiecinski1} and
Fig.\ref{fig:dsig_dpt_kwiecinski2} we show results
obtained with Kwieci\'nski UGDF \cite{Kwiecinski} 
and different combinations of factorization and 
renormalization scales as well as for different
fragmentation functions (Peterson and BCFY).
The differences between results obtained with different
combinations quantify theoretical uncertainties.
Similarly as for the standard collinear approach 
\cite{CNV05} one gets uncertainties of the order of 
a factor 2. We show individual contributions of 
electrons/positrons initiated by $c/\bar c$ or $b/\bar b$.
The contribution of the $c/\bar c$ (dashed) dominates 
at low transverse momenta of electrons/positrons.
At transverse momenta of the order of 4 - 5 GeV
the both contributions become comparable. 
We obtain rough agreement for large transverse momenta.
Similarly as for the higher-order collinear approach
\cite{CNV05} there is a missing strenght at lower
transverse momenta. A better agreement is obtained with
renormalization scale taken as transverse momentum
of the initial gluon(s). There are two strong coupling
constant in the considered order. In practice we take
$\alpha_s(k_{1t}^2) \alpha_s(k_{2t}^2)$, i.e. different
argument for each running coupling constant.
This is rather a standard prescription used in
$k_t$-factorization approach (see e.g. \cite{BS00,HKSST00})
although does not have a deep theoretical foundation.
In the latter case to avoid Landau pole we use analytic prescription
of Shirkov and Solovtsov \cite{SS97}.

\begin{figure}[!thb] 
\begin{center}
\includegraphics[width=6cm]{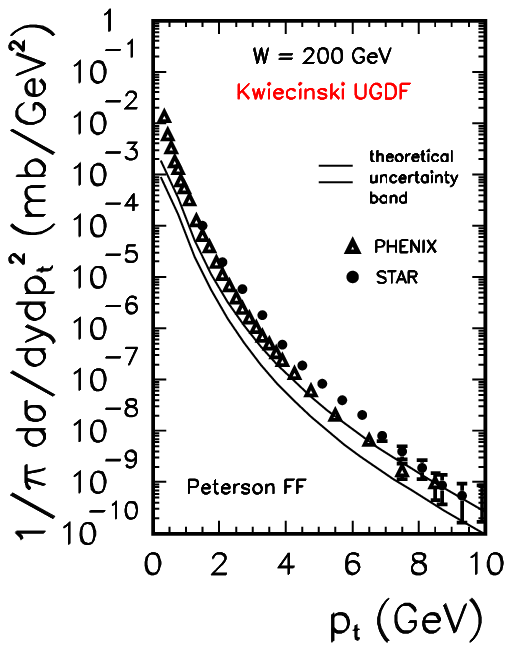}
\includegraphics[width=6cm]{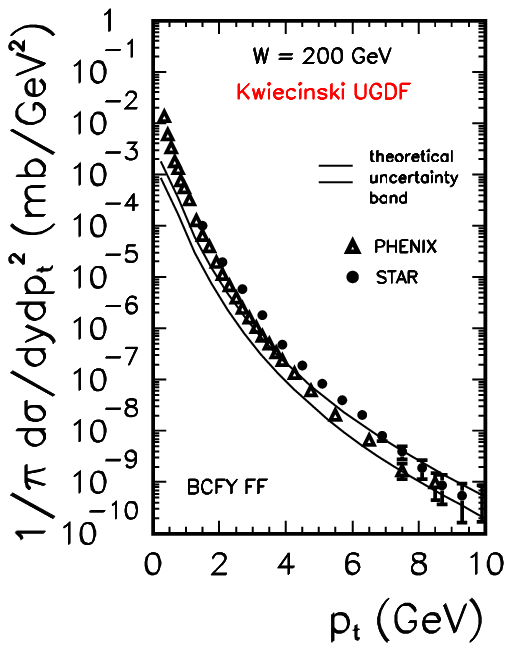}

\caption[*]{
Uncertainty band of our $k_t$-factorization calculation (both D and B decays)
with unintegrated Kwieci\'nski gluon distribution for
Peterson fragmentation function (left panel)
and BCFY fragmentation function (right panel).
The open triangles represent the PHENIX collaboration data
and the solid circles the STAR collaboration data.
\label{fig:dsig_dpt_kwiecinski_uncertainty_band}
}
\end{center}
\end{figure}

The situation for the Kwieci\'nski UGDF is summarized 
in Fig.\ref{fig:dsig_dpt_kwiecinski_uncertainty_band} where we have shown 
uncertainty band of our theoretical calculation. 
The upper curves are for
$\mu_R^2 = k_t^2$ and $\mu_F^2 = 4 m_Q^2$ and the lower
curves are for $\mu_R^2 = 4 m_Q^2$ and $\mu_F^2 = 4 m_Q^2$.
Up to now we have presented only the PHENIX collaboration
data which span a broader range of lepton transverse 
momenta.
In Fig.\ref{fig:dsig_dpt_kwiecinski_uncertainty_band} we show also the STAR
collaboration data. 
The experimental results of both groups are not
completely consistent. 
In the interval 3 GeV $<$ $p_t$(lepton) $<$ 6 GeV, the STAR data points
are somewhat higher than the PHENIX data points.
This disagreement needs further explanation.
Our results are roughly consistent with both
experimental sets at large $p_t$(lepton). 
There is a missing strenght at small transverse momenta
where only the PHENIX collaboration data exist.
This will be discussed further in the following. 

\begin{figure}[!thb] 
\begin{center}
\includegraphics[width=6.0cm]{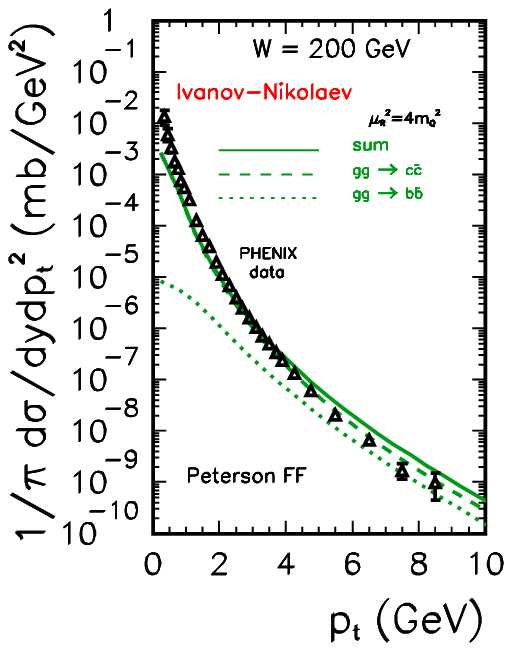}
\includegraphics[width=6.0cm]{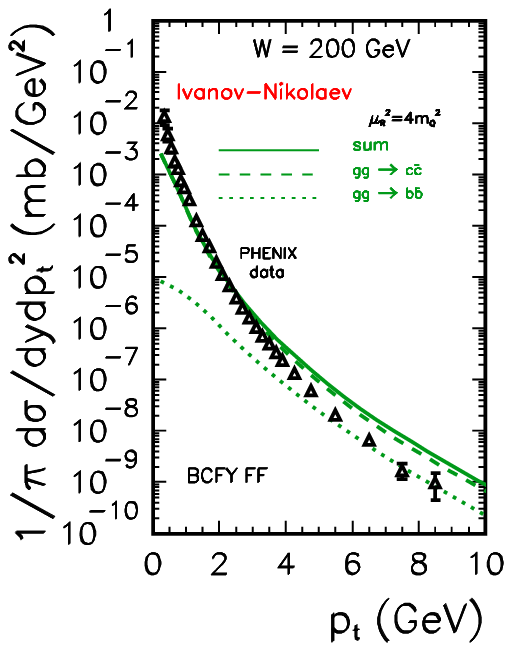}
\caption[*]{
Transverse momentum distribution of electrons/positrons
obtained with Ivanov-Nikolaev UGDF and Peterson 
(left panel) and BCFY (right panel) fragmentation 
functions.
\label{fig:dsig_dpt_ivanov_nikolaev}
}
\end{center}
\end{figure}

In Fig.\ref{fig:dsig_dpt_ivanov_nikolaev} we show
results obtained with Ivanov-Nikolaev UGDF. Although
there is some improvement at low transverse momenta,
the cross section for larger transverse momenta exceed
the experimental data.

\begin{figure}[!thb]
\begin{center}
\includegraphics[width=7.0cm]{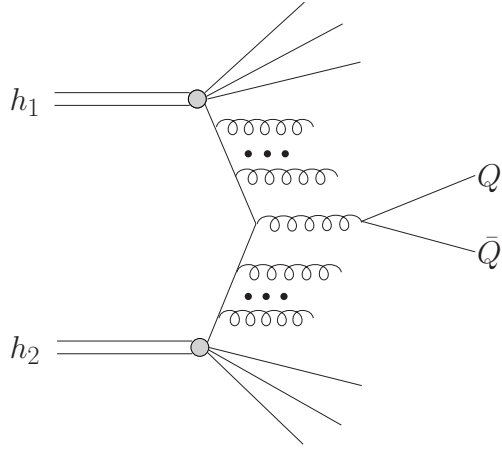}
\caption[*]{
A diagram for quark-antiquark annihilation.
\label{fig:diagram_annihilation}
}
\end{center}
\end{figure}

\begin{figure}[!thb] 
\begin{center}
\includegraphics[width=10.0cm]{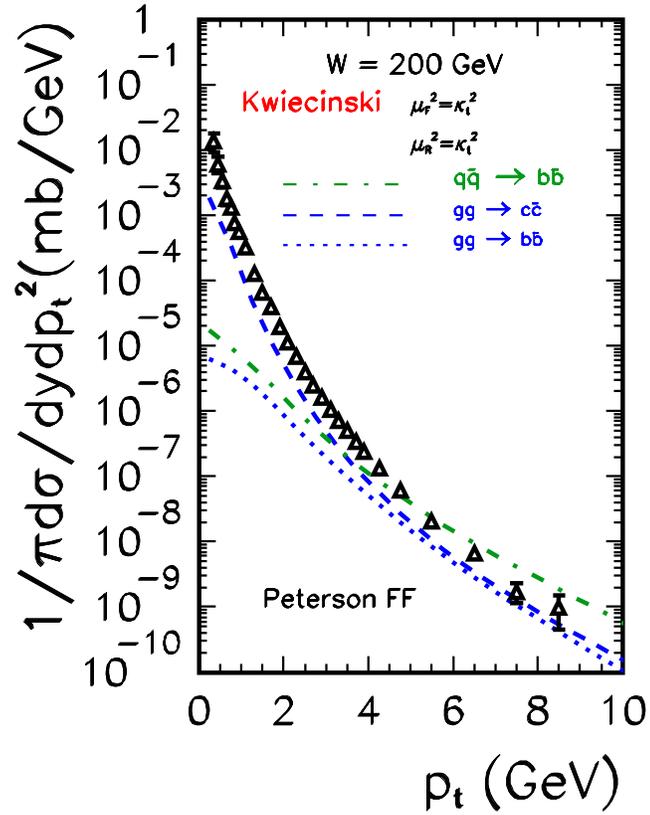}
\caption[*]{
Transverse momentum distribution of electrons/positrons
with Kwiecinski UPDFs. The dash-dotted line corresponds to
the $q \bar q \to b \bar b$ contribution.
\label{fig:dsig_dpt_kwiec_with_qqbar}
}
\end{center}
\end{figure}

It is not clear for the moment what is the missing 
strength.
Up to now we have included only gluon-gluon fusion
which is known to be dominant contribution at large
center-of-mass energies (Tevatron, LHC). 
At RHIC energies the typical longitudinal momentum 
fractions of gluons are still not too small 
$x_1, x_2 \sim$ 0.01 and the contribution of 
the quark-antiquark annihilation may be not negligible.
Therefore in the following we shall include
also quark-antiquark annihilation process.
Those processes can be included in a similar way
in the formalism of unintegrated parton distributions.
The corresponding diagram is shown in 
Fig.\ref{fig:diagram_annihilation}. 
The Kwieci\'nski formalism \cite{Kwiecinski} allows
to calculate unintegrated quark/antiquark distribution
in the same framework as unintegrated gluon distributions.
In Fig.\ref{fig:dsig_dpt_kwiec_with_qqbar} we present 
the contribution of quark-antiquark annihilation 
$q \bar q \to c \bar c$ (dash-dotted line). 
This contribution is similar in size to the 
$g g \to b \bar b$ contribution.
The contribution of $q \bar q \to b \bar b$ is negligible
and is not shown here.

\begin{figure}[!thb] 
\begin{center}
\includegraphics[width=6.0cm]{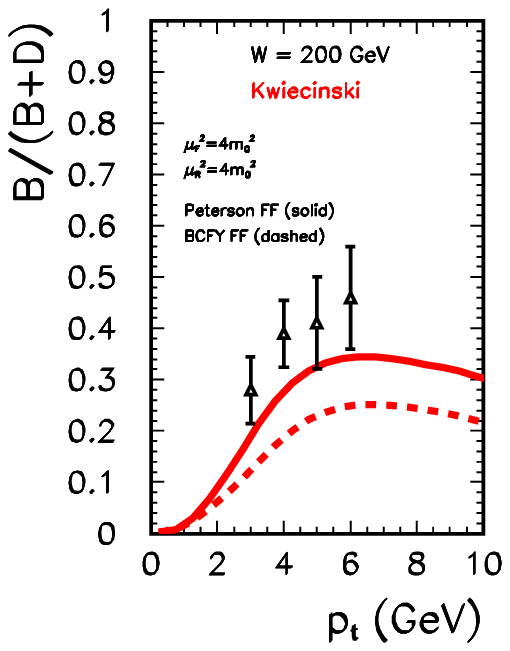}
\includegraphics[width=6.0cm]{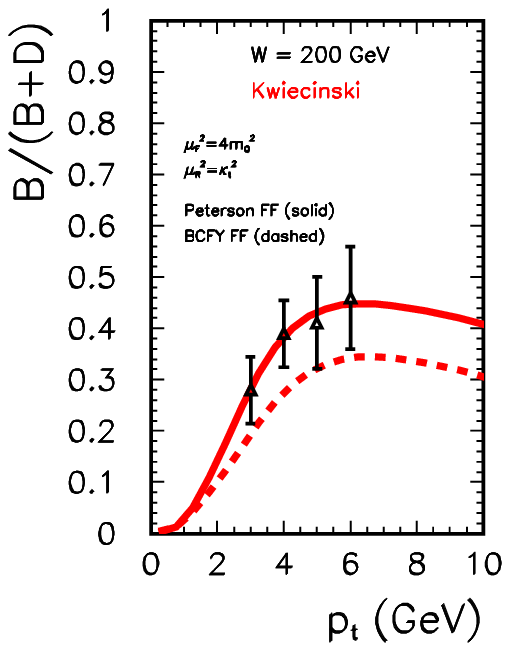}
\includegraphics[width=6.0cm]{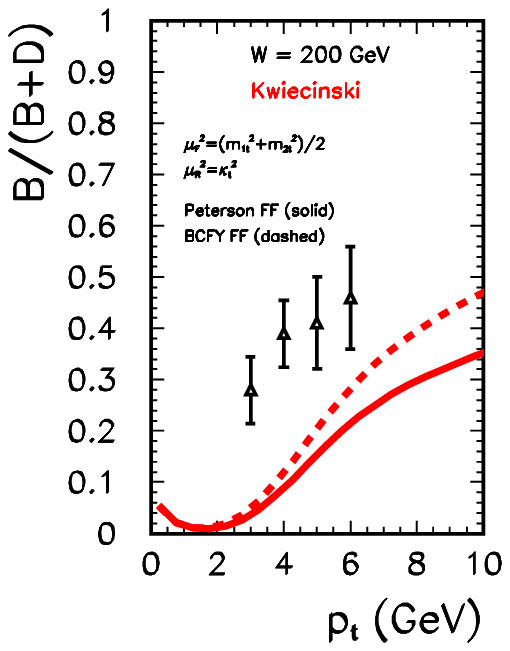}
\includegraphics[width=6.0cm]{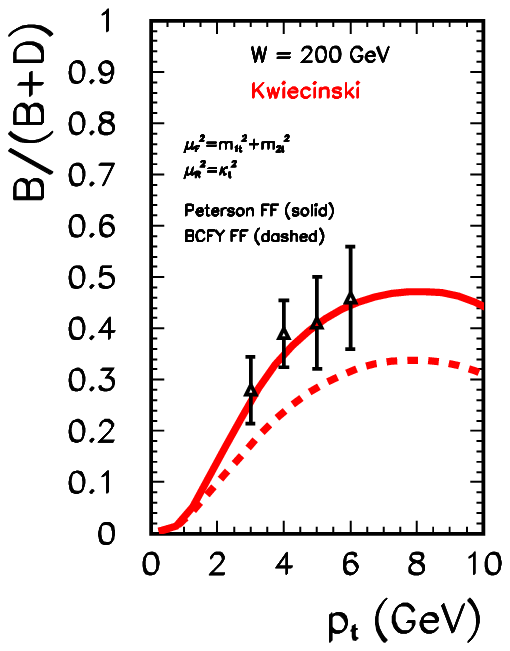}
\caption[*]{
The fraction of the B decays for the Kwieci\'nski
UGDF and different combinations of the factorization
and renormalization scales.
\label{fig:B_fraction_Kwiecinski}
}
\end{center}
\end{figure}

\begin{figure}[!thb] 
\begin{center}
\includegraphics[width=8.0cm]{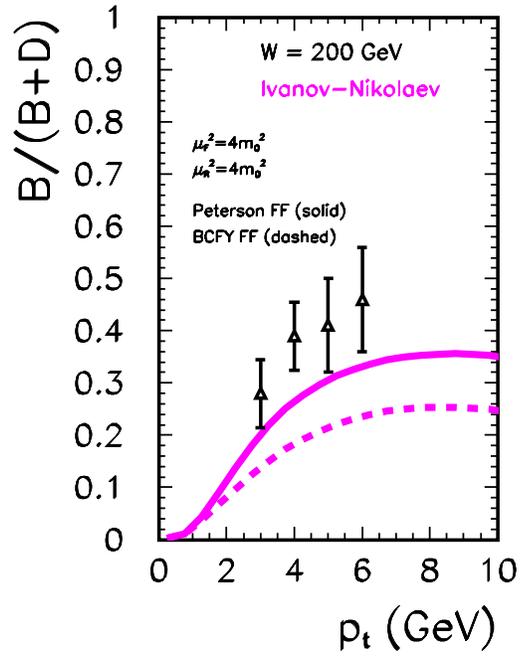}
\caption[*]{
The fraction of the B decays for the Ivanov-Nikolaev
UGDF and two different models of the fragmentation.
In this calculation $\mu_R^2 = 4 m_Q^2$.
\label{fig:B_fraction_IN}
}
\end{center}
\end{figure}

Study of nonphotonic $e^{\pm}$ and hadron correlations
allows to "extract" a fractional contribution
of the bottom mesons $B / (D + B)$ as a function
of electron/positron transverse momentum 
\cite{STAR_B_to_DB}. Recently the STAR collaboration
has extended the measurment of the relative
$B$ contribution to electron/positron transverse momenta
$\sim$ 10 GeV \cite{Mischke08a}.
In Fig.\ref{fig:B_fraction_Kwiecinski} and 
\ref{fig:B_fraction_IN}
we present our results for different unintegrated gluon
distributions and different fragmentation functions.
There is a strong dependence on the factorization
and renormalization scale in the case of the Kwieci\'nski
unintegrated gluon distributions.
A better agreement is obtained with the Peterson
fragmentation functions.
The separation into charm and bottom contributions
is very important in the context of identifying
the missing strenght. 
A new correlation method was proposed recently to identify
and separate charm and bottom production on a statistical
basis \cite{Mischke08a}. The method was tested using 
known event generators.
An alternative method of extracting the relative $B$
contribution from azimuthal angular correlations of 
nonphotonic electrons and $D_0$ mesons was proposed
\cite{Mischke08b}.
One can hope that application of the new methods
will help in disantagling the contributions
better.

\section{Discussion of the results}

We have calculated inclusive spectra of nonphotonic
electrons/positrons for RHIC energy in the framework
of the $k_t$-factorization. We have concentrated on 
the dominant gluon-gluon fusion mechanism and used
two recent unintegrated gluon distribution functions
from the literature. Special emphasis was devoted
to the Kwieci\'nski unintegrated gluon (parton)
distributions. In this formalism, using unintegrated
quark and antiquark distributions, one can calculate
in addition the quark-antiquark annihilation process
including transverse momenta of initial partons 
(quarks/antiquarks). In addition, we have used
unintegrated gluon distributions constructed by Ivanov and
Nikolaev to describe deep-inelastic data measured at HERA.

When calculating spectra of charmed ($D$, $D^*$) and 
bottom ($B$, $B^*$) mesons we have used Peterson
and Braaten et al. fragmentation functions with
model parameters from the literature. There are no big 
differences between results obtained with both 
fragmentation functions.

A very important ingredient, which influences the final
spectra, is the distribution of
electrons/positrons from the decay of $D$ and $B$ mesons.
Here we have used recent results of the CLEO and BABAR 
collaborations.
The  momentum spectra of electrons/positrons
from the decays of $D$ and $B$ mesons produced in 
the $e^+ e^-$ collisions were used in the present 
calculation to generate distribution of 
electrons/positrons coming from the decays of 
$D$ and $B$ mesons produced in the hadronic reactions.
This way we have avoided all uncertainties associated with
modeling semileptonic decays of mesons.

We have compared results obtained in our approach with
experimental data measured recently by the PHENIX 
collaboration at RHIC. We get a reasonable description 
of the data at large transverse momenta of 
electrons/positrons. Similarly as for the higher-order 
collinear approach there is a missing strength
at lower transverse momenta.

Up to now there is no clear explanation of the enhanced 
production of electrons/positrons at low transverse 
momenta.
The uncertainties related to the choice of factorization
and renormalization scale seems to be insufficient.
There can be several reasons of the 
unexplained strength at low transverse momenta.

The $k_t$-factorization approach includes many higher-order
contributions which are embodied in unintegrated gluon 
(parton) distributions. Some higher-order contributions 
are definitly not included. A simple and transparent 
example are emissions of gluons of the heavy 
quarks/antiquarks. This contribution
can be estimated in the standard collinear approach.
This effect is, however, not limited to low transverse 
momenta.

It is commonly assumed that $D/\bar D$ mesons are produced via 
fragmentation of $c/\bar c$ quarks. However, at lower 
energies (fixed target experiments) an asymmetry between 
different species of $D$ mesons have been observed 
\cite{asymmetry}. 
This asymmetry can be due to fragmentation of light 
(u,d,s) quarks/antiquarks \cite{BL06} 
($q \to D(q \bar c ) c$ or $\bar q \to D(\bar q c) \bar c$)
\footnote{There is a substantial fragmentation of light 
quarks (q $\ne$ s) in the case of kaon 
production. Such a contribution for D mesons is therefore also
not excluded.} or meson cloud 
effects \cite{CDNN01}. The asymmetry increases with 
rapidity (or Feynman $x_F$).
This makes questionable the common assumption that
$D$ mesons are produced exclusively via fragmentation
of $c$ quarks.
In this context, it would be very useful to analyze
electronic spectra at larger rapidities.
If these mechanisms are responsible for the missing
strength then the discrepancy there would be even larger.
In the moment only muons were measured at forward
rapidities \cite{PHENIX_muons} and there seems
to be some enhancement, although systematic error bars
are rather large.

The results of the PHENIX collaboration were obtained by
subtraction of several components, including decays of 
vector mesons, so-called Dalitz decays, $K_{e3}$ decays
and other mechanisms. All of them "are
concentrated" at low transverse momenta 
\cite{PHENIX_electrons}.
Only a sketch of the subtruction procedure was presented
\cite{subtraction}.
The details of the subtraction are not presented {\textit in 
extenso}.
It is therefore not clear to us how reliable such 
subtraction is.
In addition, there are several mechanisms which were not 
included. These are Drell-Yan processes, processes 
initiated by two photons
(they are expected to be concentrated at low transverse 
momenta)
and several other exclusive processes never calculated in 
the literature. It seems therefore difficult to draw 
definite conclusions before cross section for all 
these processes is evaluated. We leave such calculations 
for separate detailed studies.
In principle, also analysis of coincidence spectra, e.g. 
in invariant mass of the dilepton pair $M_{ee}$,
could help to pin down the missing mechanisms.

\vspace{1cm}

{\bf Acknowledgments}\\
We are indebted to Wolfgang Sch\"afer for useful discussion
and pointing to us some relevant references when this work
was initiated.
We are also grateful to Sergey Baranov for reading 
the manuscript and Andre Mischke for exchange of 
information on recent RHIC results and very useful 
comments.

\newpage

{\bf FIGURES}


\end{document}